\documentclass[twocolumn,pra,aps,superscriptaddress]{revtex4}

\usepackage{color}
\usepackage{epsfig}
\usepackage{latexsym}
\usepackage{amssymb}
\usepackage{amsmath}
\usepackage{algorithm}
\usepackage{algorithmic}
\usepackage{wrapfig}
\usepackage[apple mac]{inputenc}
\usepackage[english]{babel}
\usepackage{times}
\usepackage{latexsym}
\usepackage{fancyhdr}
\usepackage{verbatim}
\usepackage{tabularx}
\usepackage{epsfig}
\usepackage{amsmath}
\usepackage{amssymb}
\usepackage{graphicx}
\usepackage{wasysym}
\usepackage{hyperref}
\usepackage{tcolorbox}
\usepackage{mdframed}
\usepackage{lipsum,framed}
\usepackage{yfonts}
\usepackage{multirow}
\usepackage{booktabs}

\newcommand{\qed}{\hspace*{\fill}$\square$}

\newcommand{\be}{\begin{equation}}
\newcommand{\ee}{\end{equation}}
 \newcommand{\ket}[1]{|#1\rangle}
 \newcommand{\bra}[1]{\langle #1|}
 \newcommand{\braket}[2]{\langle #1|#2\rangle}

\begin{document}

\title{Implementing the Grover Algorithm in Homomorphic Encryption Schemes}
\author{Pablo Fern\'andez} 
\email{pabfer23@ucm.es}
\affiliation{Departamento de F\'{\i}sica Te\'orica, Universidad Complutense, 28040 Madrid, Spain.}
\author{Miguel A. Martin-Delgado}
\email{mardel@ucm.es}
\affiliation{Departamento de F\'{\i}sica Te\'orica, Universidad Complutense, 28040 Madrid, Spain.}
\affiliation{CCS-Center for Computational Simulation, Campus de Montegancedo UPM, 28660 Boadilla del Monte, Madrid, Spain.}

\begin{abstract} 
We apply quantum homomorphic encryption (QHE) schemes suitable for circuits with a polynomial number of $T/T^{\dagger}$-gates to Grover's algorithm, performing a simulation in Qiskit of a Grover circuit that contains 3 qubits. 
The $T/T^{\dagger}$ gate complexity of Grover's algorithm is also analysed in order to show that any Grover circuit can be evaluated homomorphically in an efficient manner. We discuss how to apply these QHE schemes to allow for the efficient homomorphic evaluation of any Grover circuit composed of $n$ qubits using $n-2$ extra ancilla qubits. We also show how the homomorphic evaluation of the special case where there is only one marked item can be implemented using an algorithm that makes the decryption process more efficient compared to the standard Grover algorithm.
\end{abstract}

\maketitle

\section{Introduction}
\label{sec:intro}

In recent decades, the interest in the field of quantum computation has increased greatly due to the possible benefits it could offer compared to classical computation, since the classical solution to many problems cannot be improved past a certain limit. Making use of different quantum phenomena such as superposition and entanglement, quantum algorithms, including hybrid ones, have shown significant speed-ups to many of these problems beyond what was predicted by classical computation. Some of the first examples of such algorithms are Bernstein and Vazirani's algorithm \cite{BV} or Shor's factorization algorithm \cite{Shor}. The search for more quantum algorithms and protocols continues along the developments in current quantum technologies.

Historically, the first indication of the possible advantage a quantum computer could have over a classical one came from the algorithms proposed by Deutsch \cite{Deusch1} and Deutsch-Jozsa \cite{Deuts}. In the latter work, a certain problem could be solved in one query using a quantum computer deterministically. On the other hand, this problem would require an exponential number of queries on a classical computer that used an exact model where no probability of failure was allowed. Nevertheless, if a bounded error was permitted in the algorithm, this advantage disappears \cite{bound error}.

After these algorithms, Bernstein and Vazirani created a quantum algorithm that solved a problem with a lower query complexity than what any classical computer could perform. The Bernstein-Vazirani (BV) \cite{BV} algorithm constitutes one of the first examples that showed the possible benefits a quantum computer could posses over a classical one. The most famous version of the problem is its nonrecursive version, which provided a polynomial speed-up regarding the query complexity. This query complexity was improved from the best classical algorithm that solved the problem in $O(n)$ queries to the quantum algorithm that had $O(1)$ for its query complexity. Using this algorithm as the basis they managed to construct a recursive version of the problem known as the recursive Bernstein-Varizani algorithm. This algorithm has a super polynomial speed-up compared to the best classical algorithms.

One of the most successful algorithms in quantum computing is Grover's search algorithm \cite{Grover}. This algorithm showed the advantage quantum computers have over any classical ones in the unstructured database search problem. For a database with $N$ elements the best classical algorithm has complexity $O(N)$ whereas Grover's has complexity $O(\sqrt{N})$. This constitutes a quadratic speed-up. Grover's algorithm has been studied extensively. In \cite{Grover MA} a family of quantum search algorithms is investigated, showing that Grover's algorithm holds a distinguished place in this family. The algorithm has also been used as the basis for different quantum algorithms such as the collision problem \cite{Apli Grover 1} and quantum counting \cite{quantum counting}. Surprisingly it also shares an isomorphism with classical kinematics collisions \cite{bloques}. This algorithm is one of the two building blocks of this article. The other is quantum homomorphic encryption.

In the current internet era, plenty of users take advantage of the services offered by the cloud, uploading enormous quantities of private data for many different services such as storage. Thus, ensuring the security of any computation performed on the cloud is a task of great importance. Similarly to how modern cryptography is used in order to protect the security of communications, any computations performed on the cloud should be protected using cryptographic technology that preserves the data's security. In this regard, a key technology that has been developed in recent years is homomorphic encryption. It allows a client to encrypt data and then send it to a server that performs operations on the encrypted data. Once the computations are finished, the server returns the data back so it can be decrypted. This way the client receives the evaluated data and the server never obtains any information about the actual data it  operated with. The first classical fully homomorphic encryption scheme (FHE) was created by Gentry \cite{Gentry} in 2009. Since then, more classical schemes have been proposed and perfected. These schemes are computationally secure, which means that their security is based on the difficulty of solving  mathematical problems, like ideal lattices \cite{Gentry} or the learning with errors problem (LWE) \cite{LWE}.

Regarding quantum computing, different types of technologies have been proposed to construct quantum computers, such as superconducting qubits \cite{ super1,super2,super3,super4,super5,super6, superconducting}. Currently, real quantum computers have already been developed and some of them are available in the cloud, such as the IBM Q quantum computers. Researchers have performed plenty of experiments and different types of protocols on them, like obtaining a measurement of the topological Uhlmann phase for a topological insulator \cite{uhlmann}. It is expected that most quantum computers will be accessible through the cloud in the near future. In order to preserve their security, different quantum homomorphic encryption (QHE) schemes have been developed. QHE allows a remote server to apply a quantum circuit, denoted by $QC$, on the encrypted quantum data $Enc(\rho)$ that a client has provided. Next, the client decrypts the server's output and obtains the result of the quantum circuit, $QC(\rho)$. Contrary to the classical protocols, the security of QHE schemes is based on the fundamental properties of quantum mechanics instead of the difficulty of solving complex mathematical problems.

QHE schemes can be classified as either interactive or non-interactive. If the client and server communicate during the execution of the circuit, then the scheme is said to be interactive. If interactions are allowed the efficiency of the scheme decreases since the server must wait until the client finishes some operations before it can continue executing the scheme. For this reason interactive schemes are easier to construct than non-interactive ones.
As an example of these kinds of interactive protocols, a QHE scheme where the number of interactions between client and server is the same as the number of $T$-gates contained in the evaluated circuit was proposed by Liang \cite{T interactions}.

Nevertheless, we are more concerned with studying non-interactive schemes.
Some definitions about different QHE schemes properties are given below. If a QHE scheme is $\mathcal{F}$-homomorphic then it can evaluate any quantum circuit homomorphically. A QHE scheme is said to be compact if the complexity of its decryption procedure does not depend on the evaluated circuit. Also, for a scheme to be considered a quantum fully homomorphic encryption (QFHE) scheme, it must be $\mathcal{F}$-homomorphic and compact.

The first proposals of QHE schemes began in 2012. A limited symmetric-key QHE scheme using the Boson sampling and multi-walker quantum walks was proposed by Rohde \cite{rhode}. This scheme only allows the homomorphic evaluation of certain circuits because it is not $\mathcal{F}$-homomorphic. It was later implemented experimentally by Zeuner et al. in \cite{experimento}. Liang \cite{qfhe def} defined QFHE and proposed a weak scheme that allows local quantum computations to be performed on encrypted quantum data. Nevertheless, since its evaluation algorithm depends on the secret key that the client has, its application is very limited. A QHE scheme that enables a large class of quantum computations on encrypted data was proposed by Tan \cite{Tan}. However it can not provide security in a cryptographic sense because it can only hide a constant fraction of the encrypted information.

Some researchers have also explored questions regarding theoretical limits on QHE schemes. A no-go result stating that any QFHE with perfect security must produce exponential storage overhead was proved by Yu et al. in \cite{no go result}. Constructing an efficient QFHE scheme with perfect security is impossible for this reason. An enhanced no-go result stating that constructing a QFHE scheme that is both information theoretically secure (ITS) and non-interactive is impossible was proved by Lai and Chung \cite{enhanced no go}. Thus, the best security a non-interactive QFHE scheme can achieve is computational security. Besides this no-go result, Lai and Chung constructed a compact, non-interactive ITS QHE scheme. It is not $\mathcal{F}$-homomorphic due to their no-go result so it is just a quantum somewhat homomorphic encryption (QSHE) scheme.

Due to the no-go results mentioned no QHE scheme can have non-interaction, compactness, $\mathcal{F}$-homomorphism and perfect security simultaneously. Taking this into account different schemes have been proposed with looser conditions.
For example, perfect security can be obtained if the QHE scheme is interactive, like in the one proposed by Liang \cite{T interactions} mentioned previously. Another possibility is downgrading the security level from perfect security to computational security, like Broadbent and Jeffery \cite{Broadbent} and Dulek et al. \cite{Dulek} did. Combining quantum one-time pad (QOTP) and classical FHE, two non-interactive quantum homomorphic schemes were constructed by Broadbent and Jeffery. Their security and efficiency are limited by classical homomorphic encryption schemes. Similarly, the scheme proposed by Dulek et al. is also non-interactive and depends on the construction of an ancillary gadget which is based on classical FHE. Due to this, the scheme's security is also limited to computational security.

Liang \cite{Liang} constructed two perfectly secure and non-interactive QHE schemes that are $\mathcal{F}$-homomorphic but not compact. These schemes are quasi-compact which means that the complexity of the decryption procedure scales sublinearly in the size of evaluated circuit as defined by Broadbent and Jeffery in \cite{Broadbent}. The first scheme constructed in \cite{Broadbent} is known as EPR (Einstein, Podolski and Rosen \cite{Einstein}). Its properties are quasi-compactness, $\mathcal{F}$-homomorphism, non-interaction and computational security. EPR was proved to be $M^2$-quasi-compact where $M$ is the number of $T$-gates in the evaluated circuit, which means the complexity of its decryption procedure scales with the square of the number of $T$-gates contained in the evaluated circuit. Both Broadbent's and Liang's schemes make use of Bell states and quantum measurements. However, Liang's schemes \cite{Liang} are superior regarding security since they are perfectly secure whereas EPR is only computationally secure. Furthermore one of Liang's schemes, named VGT, is $M$-quasi-compact, so it is also superior in that aspect. Since these schemes are quasi-compact, they do not contradict the no go-result mentioned previously.

The main feature of Liang's schemes \cite{Liang} is that despite being quasi-compact, they allow the efficient homomorphic evaluation of any quantum circuit with low $T/T^\dagger$-gate complexity with perfect security. For this reason they are suitable for circuits with a polynomial number of $T/T^{\dagger}$-gates. On the other hand, the decryption procedure would be inefficient for circuits that contain an exponential number of $T/T^\dagger$-gates. Gong et al. \cite{QHE  grover} have used Liang's schemes to implement a ciphertext retrieval scheme based on the Grover algorithm. They performed experiments using Qiskit and IBM's quantum computers as an example of the scheme. A quantum ciphertext dimension reduction scheme for homomorphic encrypted data based on Liang's schemes was also constructed by Gong et al. \cite{gong2}. In \cite{QHE bernstein}, Liang's schemes are applied to the recursive Bernstein-Vazirani algorithm, showing different cases in which the QHE schemes are efficient.

In this paper we propose a homomorphic implementation of Grover's algorithm using Liang's \cite{Liang} schemes. Our proposal allows the homomorphic evaluation of any Grover circuit efficiently, provided we have access to $n-2$ ancilla qubits besides the usual $n$ qubits used in a Grover search. As an example a homomorphic Grover circuit for 3 qubits has been simulated using IBM's quantum computers. Compared to Gong et al. \cite{QHE  grover}, the Grover search simulations they performed were made on a 2 qubit circuit that only needed Clifford gates. Our simulation of the homomorphic evaluation of a Grover circuit is more complex because it contains $T/T^{\dagger}$-gates. We also discuss the $T/T^{\dagger}$-gate complexity of the algorithm unlike previous works \cite{QHE  grover}. The correct states can be decrypted after the evaluation. Finally we show that the quantum search algorithm for a database with only one marked item proposed by Arunachalam and de Wolf \cite{Grover optimization}, which reduces the number of quantum gates needed to solve the problem, can be homomorphically evaluated in a more efficient manner than Grover's algorithm.

The paper is organized as follows. In section \ref{sec:application} Liang's Quantum Homomorphic Encryption schemes are reviewed. In section \ref{sec:review} Grover algorithm is reviewed. In section \ref{sec:simulation}, the simulation performed in Qiskit is shown. Then in section \ref{sec:proposal} our proposal for the general homomorphic evaluation of Grover's algorithm is discussed. Finally, the conclusions are summarized in section \ref{sec:conclusions}.

\section{Quantum homomorphic encryption scheme}
\label{sec:application}

\subsection{Preliminaries}
In the circuit model of quantum computation, quantum gates are the basic operations that constitute any circuit that can be implemented \cite{articulo MA}. The usual Clifford gates, which are $\{X,Z,H,S,CNOT\}$, will be used. Gates $X$ and $Z$ are the single-qubit Pauli gates. The Hadamard gate is $H=\frac{X+Z}{\sqrt{2}}$. Their matrix representation is given by:
\begin{equation}
X=\begin{pmatrix}
0& 1 \\
1 & 0
\end{pmatrix} \quad Z=\begin{pmatrix}
1& 0 \\
0 & -1
\end{pmatrix} \quad
H=\frac{1}{\sqrt{2}}\begin{pmatrix}
1& 1 \\
1 & -1
\end{pmatrix}.
\label{matrix1}
\end{equation}
The phase gate $S$ and $CNOT$ gate matrices are: 
\begin{equation}
S=\sqrt{Z}=\begin{pmatrix}
1& 0 \\
0 & i
\end{pmatrix}
\quad CNOT=\begin{pmatrix}
1& 0&0&0 \\
0& 1&0&0 \\
0& 0&0&1 \\
0& 0&1&0 \\
\end{pmatrix}.
\end{equation}
In order to perform universal quantum computation, a non-Clifford gate must be added to the Clifford gates, in this case it is the gate $T$. Its conjugate is simply $T^{\dagger}$. Their matrix representation is:
\begin{equation}
T= \begin{pmatrix}
1& 0 \\
0 & e^{i\frac{\pi}{4}}
\end{pmatrix} \quad
T^{\dagger}= \begin{pmatrix}
1& 0 \\
0 & e^{-i\frac{\pi}{4}}
\end{pmatrix}
\end{equation}

Then the complete set of gates is  $\mathcal{G}= \{X,Z,H,S,CNOT,T,T^{\dagger}\}$. The main issue regarding the homomorphic evaluation of a quantum circuit is evaluating the $T$ and $T^{\dagger}$-gates because these gates produce a $S$-error:
\begin{equation}
TX^aZ^b\ket{\phi}=(S^{\dagger})^aX^aZ^{a\oplus b}T\ket{\phi}.
\end{equation} 
This error has to be corrected as efficiently as possible. In Liang's schemes, it is corrected by making use of a generalization of quantum teleportation called gate teleportation.

\subsection{Gate teleportation}
\label{sec_gate_teleport}
An EPR state is an entangled quantum state given by $\ket{\Phi_{00}}=\frac{1}{\sqrt{2}} (\ket{00}+\ket{11})$. The state $\ket{\Phi_{00}}$ can be constructed by using a $H$-gate followed by a $CNOT$ acting on the state $\ket{00}$.	
From this entangled state, we can express the well known four Bell states in a compact expression:
\begin{equation}
\ket{\Phi_{ab}}=(Z^bX^a \otimes I)\ket{\Phi_{00}}, \forall a, b \in {0,1}.
\end{equation}

Quantum teleportation \cite{teleportation} is a technique that allows the transferring of quantum states between a sender and a receiver using a quantum communication channel. In this procedure, Alice wants to send Bob a qubit in the state $\ket{\psi}$, so they first share an entangled pair of qubits in the state $\ket{\Phi_{00}}$. This way both have a qubit from the EPR pair. Then Alice performs a quantum measurement in the Bell basis using the two qubits she has, $\ket{\psi}$ and one qubit of the pair $\ket{\Phi_{00}}$. Due to entanglement, Bob can now obtain the original state $\ket{\psi}$ if he applies the correct sequence of quantum gates (a combination of $X$ and $Z$) to the qubit in his possession.

For any single-qubit gate $U$, the ``$U$-rotated Bell basis'' can be defined as \cite{U rotated}: $\Phi(U)=\{\ket{\Phi(U)_{ab}}, a,b \in \{0,1\} \}$, where $\ket{\Phi(U)_{ab}}=(U^{\dagger} \otimes I)\ket{\Phi_{ab}}=(U^{\dagger}Z^bX^a \otimes I)\ket{\Phi_{00}}$. 

For a single qubit we have the following expression for quantum teleportation:
\begin{equation}
\ket{\alpha} \otimes \ket{\Phi_{00}} = \sum_{a,b\in \{0,1\}} \ket{\Phi_{ab}} \otimes X^aZ^b \ket{\alpha}.
\end{equation}

\begin{figure}[]
	\includegraphics[width=0.48\textwidth]{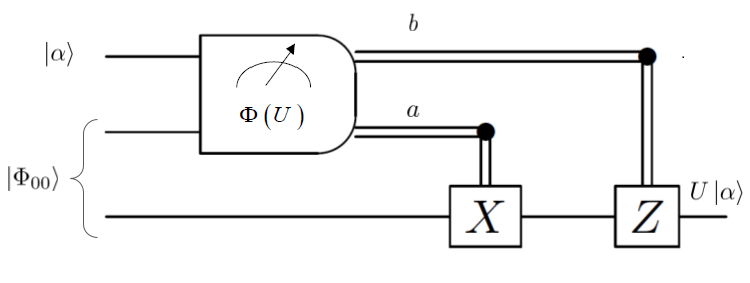}
	\centering
	\caption{Quantum circuit for gate teleportation. The box represents the quantum measurement Alice performs on the qubit $\ket{\alpha}$ and one qubit of the pair $\ket{\Phi_{00}}$. The measurement basis chosen is the $U$-rotated Bell basis $\Phi(U)$. Depending on the measurement results, $a$ and $b$, Bob applies $X^{a}$ and $Z^{b}$ in order to obtain $U \ket{\alpha}$.
	}
	\label{gatetelepory}
\end{figure}

It can easily be extended for the ``$U$-rotated Bell basis'':
\begin{equation}
\ket{\alpha} \otimes \ket{\Phi_{00}} = \sum_{a,b\in \{0,1\}} \ket{\Phi(U)_{ab}} \otimes X^aZ^bU \ket{\alpha}.
\label{gate teleport}
\end{equation}
where $U$ is any single-qubit gate. Then equation $\eqref{gate teleport}$ describes gate teleportation. Just like in the usual quantum teleportation, Alice and Bob first share an entangled EPR pair in the state $\ket{\Phi_{00}}$. Then Alice prepares a qubit in the state $\ket{\alpha}$ and performs a ``$U$-rotated Bell measurement''. This is simply a quantum measurement in which the $U$-rotated Bell basis is selected as its measurement basis. She performs the measurement on the two qubits she possesses, one in the state $\ket{\alpha}$ and one of the EPR pair in the state $\ket{\Phi_{00}}$. From this measurement she obtains the results $a$ and $b$. After the measurement Bob's qubit transforms into the state $X^aZ^bU \ket{\alpha}$. Next, Alice tells Bob the results of her measurement, so Bob can apply the correct Pauli $X$ and $Z$ operators to finally obtain $U \ket{\alpha}$. The gate teleportation procedure is represented in figure \ref{gatetelepory}.

Gate teleportation is a clear extension of quantum teleportation since if $U$ were the identity then the $U$-rotated Bell basis $\Phi(U)$ would reduce to the standard Bell basis, and therefore gate teleportation would implement the standard quantum teleportation protocol.

This gate teleportation protocol is the basis of Liang's QHE schemes, because it corrects the error that results from the homomorphic evaluation of a $T$-gate.

\subsection{Encryption, evaluation and decryption}
\label{sec:encryption}
Suppose that there is a quantum circuit composed of gates from the set $\mathcal{G}= \{X,Z,H,S,CNOT,T,T^{\dagger}\}$ that acts on $n$ qubits. There are $l$ gates in the circuit in total. The quantum gates in the circuit are numbered starting from the left to right. This way the first quantum gate is denoted by $G[1]$, the second by $G[2]$ and so on until the last gate $G[l]$. The $j$th quantum gate is denoted by $G[j]$. The qubit it acts on is denoted by the subscript $w$.
For example the gate $G[j]=X_w$ is a Pauli $X$-gate acting on the $w$th qubit of the circuit. The $CNOT$ gate acting on the $w$th control qubit and the $w'$th target qubit is denoted by $CNOT_{w,w'}$. Among the total number of gates $l$, the total number of $T$ and $T^{\dagger}$ gates are denoted by $M$. Each $T$ and $T^{\dagger}$ gate has its own number, $j_i$ ($i\leq j_i\leq l$, $1\leq i \leq M$) in the sequence of gates \{$G[j], j=1,2,..,l$\}. Then $G[j_i]=T/T^{\dagger}$ where $j_i<j_{i+1}$. If $G[j_i]$ ($1 \leq i \leq M$) is applied on the $w_i$th qubit ($1 \leq w_i \leq n$) then we can write $G[j_i]=T_{w_i}/T_{w_i}^{\dagger}$.

The first step in Liang's schemes is encrypting the data that will be sent to the server. This is achieved by applying a symmetric-key encryption scheme, called quantum one time pad (QOTP), to the data. It consists on applying a combination of $X^a$ and $Z^b$ gates to each qubit. The bits $a$ and $b$ are randomly selected from $\{0,1\}$ and constitute the secret key of the client $sk$. If the plaintext data contains $n$ qubits, the secret key $sk$ has $2n$ bits $sk=(a_0,b_0),a_0,b_0\in\{0,1\}^n$. The $w$th plaintext qubit is encrypted using the secret bits $(a_0(w), b_0(w))$ such that: $\ket{\alpha}_w \rightarrow X^{a_0(w)}Z^{b_0(w)} \ket{\alpha}_w=\ket{\rho_0}_w$ where $\ket{\rho_0}_w$ represents the encrypted state of the $w$th qubit. 

QOTP was proposed by Boykin and Roychowdhury \cite{one time pad}. They proved that if the bits $a$ and $b$ are randomly selected from $\{0,1\}$ and used only once, QOTP has perfect security.
This is because if QOTP is applied to any arbitrary quantum state $\sigma$, the totally mixed state  $\frac{I_{2^n}}{2^n}$ is obtained: $\frac{1}{2^{2n}}\sum_{a,b\in \{0,1\}^n} X^aZ^b\sigma(X^aZ^b)^{\dagger}=\frac{I_{2^n}}{2^n}.$

\begin{figure}[]
	\includegraphics[width=0.5\textwidth]{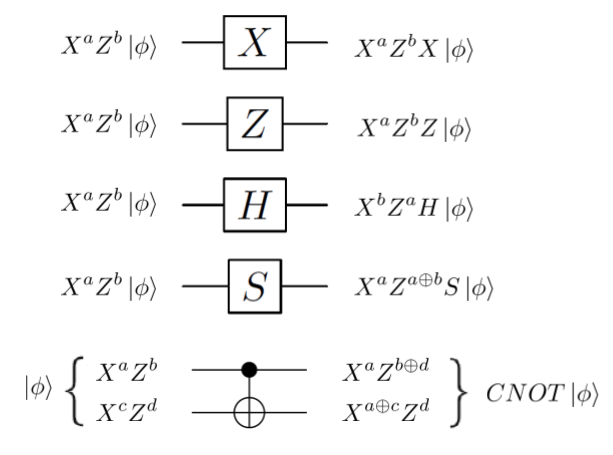}
	\centering
	\caption{Key updating rules for the homomorphic evaluation of Clifford gates.
	}
	\label{reglas}
\end{figure}

Next, the encrypted data will be sent to the server. In order to complete its designated quantum circuit, the server performs quantum gates from the set $\mathcal{G}= \{X,Z,H,S,CNOT,T,T^{\dagger}\}$. As long as the gates are Clifford and not $T$ or $T^{\dagger}$, the homomorphic evaluation can be performed easily. Each time the server performs one of these gates on a qubit, the key is updated according to the algorithm shown in figure \ref{reglas}. After the $j$th ($1\leq j \leq l-1$) gate is performed, a new key (denoted as
$(a_j, b_j), a_j , b_j \in\{0,1\}^n$) is obtained through key updating. This is the intermediate key. As an example, the new key that would be obtained after applying a $H$-gate is $(a_1,b_1)=(b_0,a_0)$ assuming it was the first gate in the circuit so $j=1$.

\begin{figure}[]
	\includegraphics[width=0.52\textwidth]{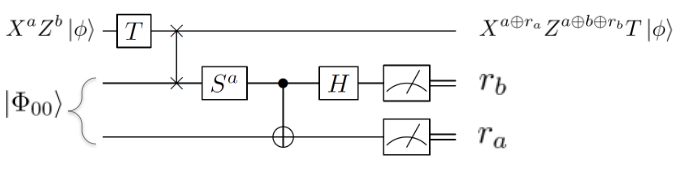}
	\caption{Homomorphic evaluation of $T$-gate. $\ket{\Phi_{00}}$ represents an EPR pair. 
	}
	\label{T gate}
\end{figure}

If the gate is a $T/T^{\dagger}$, its homomorphic evaluation is performed using the gate teleportation procedure explained in section \ref{sec_gate_teleport}. In this case, the $S^a$-rotated Bell measurement is used in order to correct the $S$-error described previously. At the start of the circuit's evaluation an EPR source generates $M$ Bell states, as many as $T/T^{\dagger}$-gates are in the circuit, denoted by $\{\ket{\Phi_{00}}_{c_i,s_i}, i=1,..,M\}$, where qubits $c_i, i = 1,....,M$ and qubits $s_i, i =
1,....,M$ are kept by the client and server respectively. If the server has to evaluate a $T/T^{\dagger}$-gate, it first applies the gate to the desired qubit, then performs a SWAP gate between this encrypted qubit and one of the entangled qubits the server possesses, $s_i$. The next step is applying a $S^a$-rotated Bell measurement on the qubits $s_i$ and $c_i$. Using the results from this measurement, $r_a$ and $r_b$, the encryption key can be updated. From $TX^aZ^b\ket{\alpha}$, this whole process returns $X^{a\oplus r_a}Z^{a \oplus b \oplus r_b}T\ket{\alpha}$. In case the evaluated gate is a $T^{\dagger}$-gate, the key will be updated from $T^{\dagger}X^aZ^b\ket{\alpha}$ to $X^{a\oplus r_a}Z^{ b \oplus r_b}T^{\dagger}\ket{\alpha}$. This whole process is shown in figure \ref{T gate}. It will be performed $M$ times, as many as $T/T^{\dagger}$-gates are in the circuit.

The $M$ measurements can be postponed until the server has finished all the quantum operations due to the principle of deferred measurement. The server must also generate the key-updating functions according to the key-updating rules of each gate from the set $\mathcal{G}$ that have already been explained. For the scheme VGT, the server must generate $2M+1$ key-updating functions $\{g_i\}_{i=1}^{M}$ and $\{f_i\}_{i=1}^{M+1}$. Once the quantum circuit has been completed, the server sends all the encrypted qubits, the key-updating functions based on the key-updating rules and all the ancillary $s_i$ qubits to the client. Then the client updates its keys according to the key-updating functions and performs a $S^a$-rotated Bell measurement on qubits $c_i$ and $s_i$. Since to perform the correct $S^a$-rotated Bell measurement the updated $a$ key is needed, the client needs to alternate between updating the keys and measuring. These measurements must be performed in the pre-established order, since the updated key depends on the result of the previous measurement. Once all measurements are completed and all the keys updated, the client obtains the final key $dk=(a_{\text{final}},b_{\text{final}}), a_{\text{final}}, b_{\text{final}}\in\{0,1\}^n$. Finally the client decrypts the qubits that are in the state $\ket{\rho_\text{\text{final}}}$, which is the final state that the server outputs, using the final key to obtain the plaintext results in the state $\ket{\alpha_{\text{final}}}$: $X^{ a_{\text{final}}}Z^{b_{\text{final}}}\ket{\rho_{\text{final}}}=\ket{\alpha_{\text{final}}}$. 

The whole QHE scheme can be described in five steps: \textbf{Setup}, \textbf{Key Generation}, \textbf{Encryption}, \textbf{Evaluation}, \textbf{Decryption}. 
\begin{enumerate}
	\item \textbf{Setup}: an EPR source generates $M$ Bell states, as many as $T/T^{\dagger}$-gates are in the circuit, $\{\ket{\Phi_{00}}_{c_i,s_i}, i=1,\dots,M\}$, where qubits $c_i, i = 1,\dots,M$ and qubits $s_i, i =1,\dots,M$ are kept by the client and the server respectively.
	\item \textbf{Key Generation}: Generate random bits $a_0$, $b_0$ $\in\{0,1\}^n$ and output the secret key $sk=(a_0,b_0)$.
	\item \textbf{Encryption}: for any $n$ qubit data $\ket{\alpha}$, client performs
	QOTP encryption using the secret key $sk = (a_0, b_0)$: $\ket{\alpha} \rightarrow X^{a_0}Z^{b_0} \ket{\alpha}=\ket{\rho_0}$.
	\item  \textbf{Evaluation}: The server applies the quantum gates $G[1]$, $G[2]$,\dots, $G[l]$ in order on the $n$ encrypted qubits. For each $j\in\{1,\dots,l\}$ there are two possible cases: if $j \notin \{j_1,\dots,j_M\}$ then $G[j]$ is a Clifford gate and the server applies it. If $j=j_i$ $(1,\leq i \leq M)$ then the gate $G[j]=G[j_i]=T_{w_i}$ or $T^{\dagger}_{w_i}$, the server performs this gate $G[j]$ on the qubit $w_i$ and then the server applies a SWAP gate on this $w_i$th qubit and one of the entangled qubits $s_i$.
	Taking $j_0=0$ and using the key-updating rules the server generates the polynomial $\{g_i\}_{i=1}^{M}$ for one key bit $a_{j_i-1}(w_i)=g_i(a_{j_{i-1}},b_{j_{i-1}})$, $i=1,\dots,M$, with $a_{j_i-1}\in\{0,1\}$. Likewise according to the key updating rules the server generates the polynomial $\{f_i\}_{i=1}^{M}$ for the intermediate key $(a_{j_i},b_{j_i})=f_i(a_{j_{i-1}},b_{j_{i-1}},r_a(i),r_b(i))$, $i=1,...,M$, with $(a_{j_{i}},b_{j_{i}})\in\{0,1\}^{2n}$. The final polynomial that the server generates is $f_{M+1}$ for the final key $(a_{\text{final}}, b_{\text{final}})=f_{M+1}(a_{j_M}, b_{j_M})$, with $(a_{\text{final}}, b_{\text{final}})$ $\in\{0,1\}^{2n}$. After the last gate is applied the server sends all the encrypted qubits, the key-updating functions and all the ancillary $s_i$ qubits to the client.
	\item  \textbf{Decryption}: The client alternates between key updating and measurements. For each $i=1,\dots,M$, the client computes $g_i$ and obtains the corresponding $a$ for the $i$th $S^a$-rotated Bell measurement: according to the key $(a_{j_{i-1}},b_{j_{i-1}})$ and the key-updating function $g_i$, the client obtains $a=g_i(a_{j_{i-1}},b_{j_{i-1}})$. If $i=1$, the secret key $sk=(a_0,b_0)$ is $(a_{j_{i-1}},b_{j_{i-1}})$. Then using this measurement basis the client performs a $S^a$-rotated Bell measurement on qubits $c_i$ and $s_i$ and obtains the measurement results $r_a(i)$ and $r_b(i)$. Then the client computes the intermediate key $(a_{j_{i}},b_{j_{i}})$ according to the key updating function $f_i$: $(a_{j_i},b_{j_i})=f_i(a_{j_{i-1}},b_{j_{i-1}},r_a(i),r_b(i))$. After the last round, $i=M$, of this process is performed the client obtains the intermediate key $(a_{j_M},b_{j_M})$. Using this key and the last key-updating function $f_{M+1}$ the client obtains the final key: $(a_{\text{final}}, b_{\text{final}})$. Finally the client performs QOTP decryption on the encrypted qubits to obtain the final states: $X^{a_{\text{final}}}Z^{b_{\text{final}}}\ket{\rho_{\text{final}}}=\ket{\alpha_{\text{final}}}$.
	
\end{enumerate}

A remark about the setup step in which the client and server share $M$ Bell states is in order. It is not really necessary for them to pre-share the Bell states. Instead, a Bell state can be generated by the server each time it evaluates a $T/T^{\dagger}$-gate. Then the server can produce the $M$ Bell states required and send them back to the client along with all the encrypted qubits and the key-updating functions.

Therefore, this scheme can evaluate any quantum circuit homomorphically ($\mathcal{F}$-homomorphic) with perfect security. The server can never learn any information about the plaintext or the evaluation keys at any point. The data the server receives is encrypted with QOTP so it is perfectly secure and the secret key $sk$ is hidden perfectly too. There are no interactions in the evaluation process so the server can not obtain any information about the plaintext there either. Finally once the data is sent back to the client for the decryption process it is impossible to obtain any information, since the client performs quantum measurements and key-updating locally and does not interact with the server. 
Now that the whole scheme has been explained, the reason why it is quasi-compact becomes clear. Unlike Clifford gates, the number of $T/T^{\dagger}$-gates contained in the evaluated circuit make the complexity of the decryption process grow due to the quantum measurements the client has to perform in succession. For this reason the discussed scheme is only suitable for circuits with a polynomial number of $T/T^{\dagger}$-gates.

\section{Grover's algorithm review}
\label{sec:review}
In the unstructured search problem, a list of $N$ unordered elements is given. The components of this list are labelled from 0 to $N-1$ without any loss of generality. There is an element in this list that has to be found called $w$. In the classical scenario, there is no structure in this list so all the elements have to be checked in order to find the desired element. This takes on average $N/2$ attempts and at worst it takes $N$ tries, making the complexity of the classical problem $O(N)$. 

In the quantum algorithm given by Grover, the searched element can be found in just $O(\sqrt{N})$ steps which constitutes a quadratic speedup. For simplicity, the elements of the list can be written as $N=2^n$ for some integer $n$. This way the $N$ elements of the list can be represented using $n$ qubits. 

The algorithm begins with $n$ qubits in the $\ket{0}$ state. The first step in Grover's algorithm is state preparation. In this step, $n$ Hadamard gates are applied to each qubit in order to obtain an uniform superposition of all possible $n$ bit strings:
\begin{equation}
\ket{s}=H^{\otimes n} \ket{0^{\otimes n}}=\frac{1}{\sqrt{N}}\sum_{x=0}^{N-1}\ket{x}.
\end{equation}
In the next step, the oracle denoted by $U_w=I-2\ket{w}\bra{w}$ is applied. 
Given any state $\ket{x}$ as input, the oracle will output:
\begin{equation}
U_{w}\ket{x} =
\begin{cases}
\ket{x} & \text{if $x\neq w$}\\
-\ket{x} & \text{if $x=w$}\\
\end{cases}    
\label{eq1}   
\end{equation}
This oracle changes the amplitude of the desired state while leaving the rest unaffected. The action of the oracle on $\ket{s}$ is given by:
\begin{equation}
U_{w}\ket{s} = \frac{1}{\sqrt{N}}\sum_{\substack{x=0\\x\neq w}}^{N-1}\ket{x}-\frac{1}{\sqrt{N}}\ket{w}. 
\label{eq2}    
\end{equation}
The next step of the algorithm is applying the diffusion operator given by $U_s=2\ket{s}\bra{s}-I$ to $U_w\ket{s}$. This operator flips the amplitudes around the mean. This makes the amplitude of each state decrease except in the case of desired state because its amplitude increases. This process of amplifying the desired state's amplitude is known as amplitude amplification. Grover's algorithm consists on applying the operators $U_w$ and $U_s$ iteratively. 

\begin{figure}[]
	\includegraphics[width=0.25\textwidth]{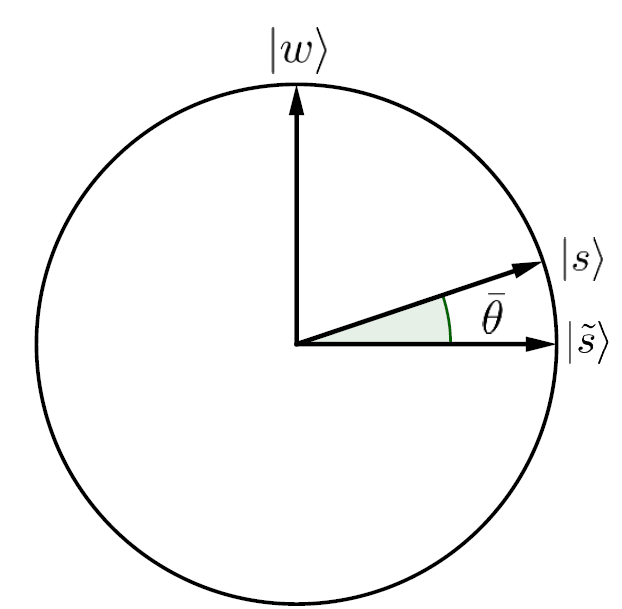}
	\centering
	\caption{Circle defined by eq. \eqref{eq3}. The basis is $\ket{\tilde{s}}$ and $\ket{w}$.
	}
	\label{circulo}
\end{figure}

The algorithm has a well known geometric interpretation based on two reflections that lead to a rotation around an angle $\theta$ in a plane. In this interpretation, an orthonormal coordinate system is used. Since the state $\ket{s}$ is not orthogonal to $\ket{w}$ we can introduce a new state $\ket{\tilde{s}}$ that is perpendicular to $\ket{w}$ defined by:
\begin{equation}
\ket{\tilde{s}}=\frac{1}{\sqrt{N-1}}\sum_{\substack{x=0\\x\neq w}}^{N-1}\ket{x}.
\end{equation}
$\ket{\tilde{s}}$ is then obtained from $\ket{s}$ by removing the desired state $\ket{w}$ and rescaling so that the state $\ket{\tilde{s}}$ still has its norm equal to 1. The states $\ket{\tilde{s}}$ and $\ket{w}$ form a basis and any state can be expressed in terms of an angle $\theta$:
\begin{equation}
\ket{\theta}=\cos(\theta)\ket{\tilde{s}}+\sin(\theta)\ket{w}.
\label{eq3}
\end{equation}
Both $U_w$ and $U_s$ keep the resulting state in the circle defined by eq \eqref{eq3}. This circle is represented in figure \ref{circulo}. The first step of the algorithm is also represented in figure \ref{circulo} since the state shown is the starting superposition state $\ket{s}$. The angle $\bar{\theta}$ that this state forms with the horizontal axis is given by equation \eqref{angulo}.
\begin{equation}
\bar{\theta}=\arcsin\braket{s}{w}=\arcsin\frac{1}{\sqrt{N}}.
\label{angulo}
\end{equation}

\begin{figure}[]
	\includegraphics[width=0.35\textwidth]{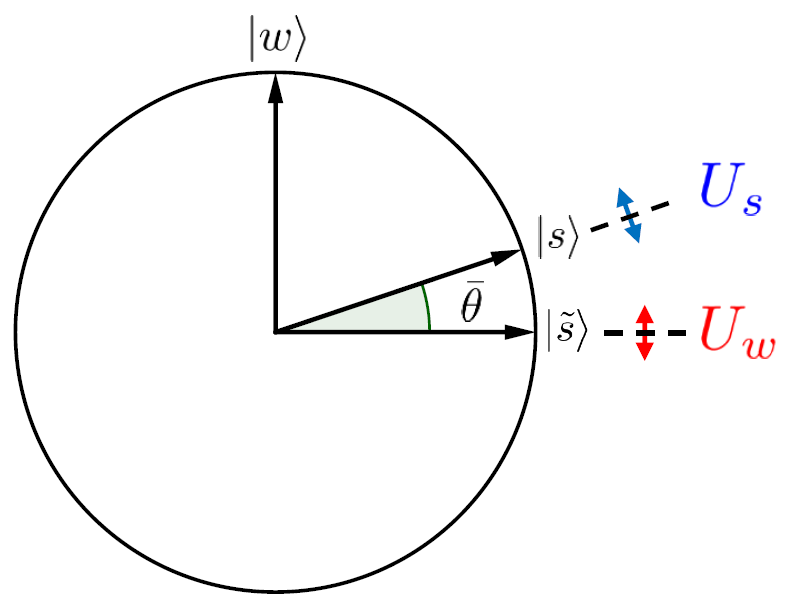}
	\centering
	\caption{$U_w$ reflects about $\ket{\tilde{s}}$ and $U_s$ reflects around $\ket{s}$.
	}
	\label{circulo2}
\end{figure}

The next step is applying the oracle $U_w$. Geometrically this is equivalent to a reflection about the $\ket{\tilde{s}}$ axis since the amplitude of $\ket{w}$ changes its sign. The last step to complete an iteration is applying $U_s$. This operator reflects the state about $\ket{s}$ as seen in figure \ref{circulo2}. 

\begin{figure}[]
\includegraphics[width=0.25\textwidth]{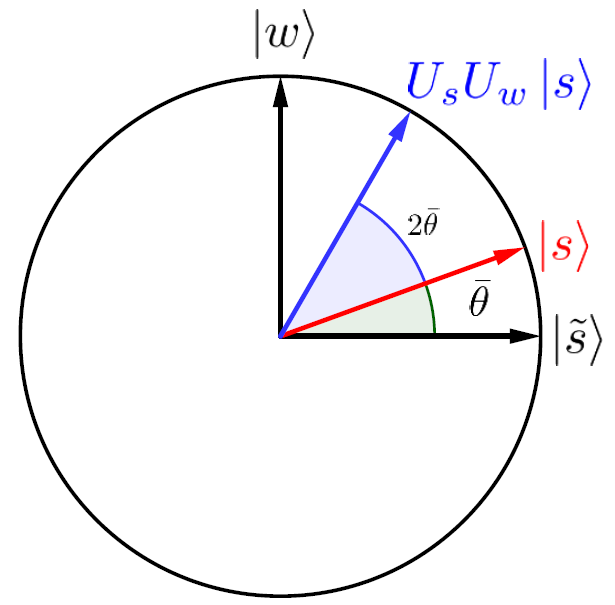}
\centering
\caption{The combination of $U_w$ and $U_s$ rotate the initial state $2\bar{\theta}$.
}
\label{circulo3}
\end{figure}

Two consecutive reflections constitute a rotation. The angle corresponding to this rotation is $2\bar{\theta}$ as seen in figure \ref{circulo3}. This way the combined action of both operators on any state is given by eq. \eqref{rotacion}. 
\begin{equation}
U_sU_w\ket{\theta}=\ket{\theta+2\bar{\theta}}.
\label{rotacion}
\end{equation} 
Applying both operators to the initial state $\ket{s}$ gives $U_sU_w\ket{s}=\ket{\bar{\theta}+2\bar{\theta}}$.

The combination of the operators is then repeated iteratively in order to rotate the initial $\ket{s}$ state closer to the state $\ket{w}$. This means rotating from $\theta=\bar{\theta}$ to $\theta=\pi/2$ in steps of $2\bar{\theta}$. After $O(\sqrt{N})$ iterations, the amplitude of the state $\ket{w}$ reaches its maximum so the probability of obtaining $\ket{w}$ in a quantum measurement approaches $1$. The final step of the algorithm is then a quantum measurement of the $n$ qubits that returns the desired element of the list $w$ with high probability.

In case there are $m$ elements to be found in the list instead of just one, the steps of the algorithm remain the same: preparation of the $\ket{s}$ state, application of the operators $U_w$ and $U_s$ iteratively and measurement of the $n$ qubits. It can be shown that the number of steps required to obtain the desired elements is bounded by $O\left(\sqrt{\frac{N}{m}}\right)$ \cite{Grover varias sols} in this case.

\section{Homomorphic Grover simulation in Qiskit}
\label{sec:simulation}
In this section a simulation of a homomorphic evaluation of a Grover circuit is shown. The simulation has been performed in IBM Qiskit \cite{qiskit}.

The client wants to solve the unstructured search problem using Grover's algorithm. It wants to do it without the server learning anything about the data so it decides to use quantum homomorphic encryption to preserve its security. 

\begin{figure}[]
	\includegraphics[width=0.45\textwidth]{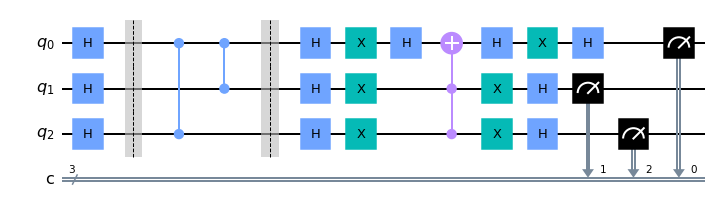}
	\centering
	\caption{The quantum circuit that the client wants to evaluate. It implements a Grover search for $n=3$ qubits, $N=8$ elements and $m=2$ marked items in Qiskit.
	}
	\label{circuito_base}
\end{figure}

The circuit that the client wants to evaluate is shown
in figure \ref{circuito_base}. The qubits are denoted by $q_0$, $q_1$, ..., $q_{n-1}$. Notice that in Qiskit the states are represented as $\ket{q_{n-1}...q_1q_0}$ so the circuit in figure \ref{circuito_base} should be read starting from $q_2$ and finishing at $q_0$. The length of the list is $N=8$ and so $n=\log N=3$ qubits are needed. This circuit finds two elements in the list: the states $\ket{011}$ and $\ket{101}$ (recall that Qiskit notation is being used here). Then Grover's algorithm for two marked elements is applied. The reason why this circuit was selected is that for $N=8$ and $m=2$ only one iteration is needed and the circuit still requires the use of $T/T^{\dagger}$-gates. This makes the homomorphic evaluation not trivial but still simple enough to be simulated.

\begin{figure}[]
	\includegraphics[width=0.45\textwidth]{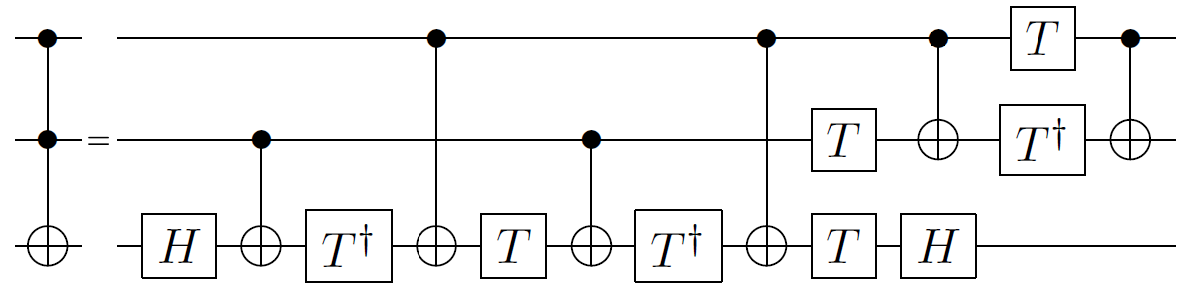}
	\centering
	\caption{A Toffoli gate can be decomposed into 7 $T/T^{\dagger}$-gates.
	}
	\label{Toffolis}
\end{figure}

All the steps of the circuit are separated by a grey barrier in the figure. The first step applies 3 $H$-gates to each qubit. The next step is the application of the oracle $U_w$, so the states $\ket{011}$ and $\ket{101}$ are marked using two controlled-$Z$ gates. The third step is the application of the diffusion operator $U_s$ using a $CCZ$ gate (a controlled $Z$-gate that has 2 control qubits and one target qubit) surrounded by $H$ and $X$ gates. In figure \ref{circuito_base}, the $CCZ$ gate has already been decomposed into a Toffoli gate which has its target qubit surrounded by two $H$-gates. Since $Z=HXH$, any $Z$-gate can be implemented using $H$, $X$ and $H$.    
The reason for this decomposition is that in order to implement this circuit homomorphically, it has to be decomposed into gates of the set $\mathcal{G}= \{X,Z,H,S,CNOT,T,T^{\dagger}\}$. This means that each controlled-$Z$ gate from $U_w$ has to be substituted by a $CNOT$ gate surrounded by two $H$-gates just like the $CCZ$ gate was substituted by a Toffoli gate surrounded by $H$-gates. The last gate that has to be transformed is the Toffoli gate which can be decomposed into 7 $T/T^{\dagger}$-gates as seen in figure \ref{Toffolis}.

\begin{figure}[]
	\includegraphics[width=0.45\textwidth]{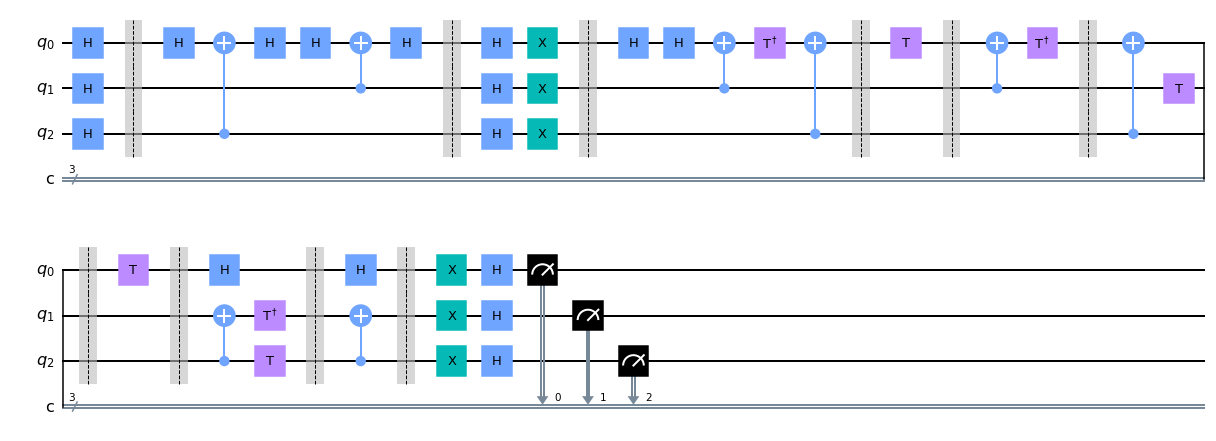}
	\centering
	\caption{Compilation of the circuit that the client wants to evaluate decomposed using gates from the set $\mathcal{G}$.
	}
	\label{circuito_evaluar}
\end{figure}

After decomposing the two controlled-$Z$ gate, the $CCZ$ gate and the Toffoli gate into gates of the set $\mathcal{G}$, a new circuit represented in figure \ref{circuito_evaluar} is obtained. This is the circuit that has to be evaluated homomorphically using the QHE scheme described in section \ref{sec:application}. The total number of gates of the circuit is $l=35$. The sequence of gates is $G[1]=H_0$, $G[2]=CNOT_{2,0}$, ... and so on until the last three gates which are $G[33]=H_0$, $G[34]=H_1$, $G[35]=H_2$.

\begin{figure*}[]
	\includegraphics[width=0.8\textwidth]{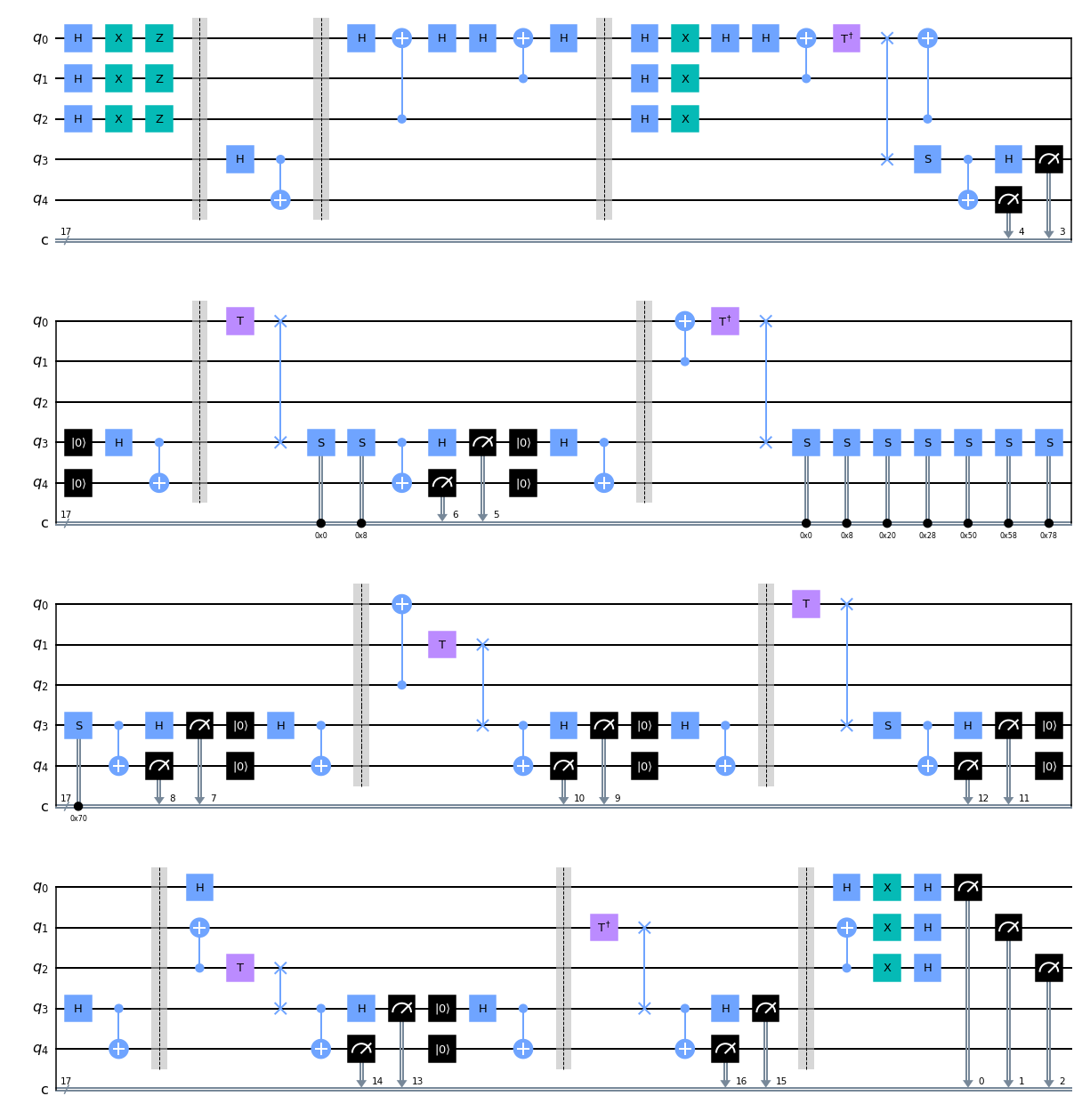}
	\centering
	\caption{Simulated circuit in Qiskit. This circuit is a simplified version of the simulation performed in order to make the image more compact.
	}
	\label{circuito_final}
\end{figure*}

Following the QHE scheme explained the client starts by generating the random bits $a_0$, $b_0$$\in\{0,1\}^3$ to obtain his secret key $sk=(a_0,b_0)$. In the simulation $n=3$. We chose $a_0=a_0(0)a_0(1)a_0(2)=111$ and $b_0=b_0(0)b_0(1)b_0(2)=111$ where $a_0(n-1)$ and $b_0(n-1)$ refer to the values of the key for $q_{n-1}$. This way we have that for $q_0$ the keys are $(a_0(0), b_0(0))=(1,1)$. For $q_1$ and $q_2$ the keys are $(a_0(1), b_0(1))=(1,1)$ and $(a_0(2), b_0(2))=(1,1)$ respectively. Next, the first step of Grover's algorithm (state preparation) is performed using 3 $H$-gates to obtain the superposition state $\ket{s}$. Then $\ket{s}$ is encrypted using the secret key $sk$. This means that all the qubits are encrypted using $X^1Z^1$ since the keys for each qubit are $(1,1)$. Now that the data has been encrypted by the client, it is sent to the server. 

The server starts by generating as many EPR pairs as $T/T^\dagger$ are in the circuit. Since $M=7$ then 7 entangled pairs have to be generated. An EPR pair can be easily constructed just using a $H$-gate on the first qubit and a $CNOT$ gate in which the target is the second qubit. 14 additional qubits are then used in the simulation. 

The simulation we performed contains 17 qubits in total. An image containing such a number of qubits would be too large to show properly so we show an equivalent circuit to the one that we used in our Qiskit simulation in figure \ref{circuito_final}. Besides the Qiskit simulation we also calculated the value of each key before any $S^a$-rotated Bell measurement is performed.
There are some details that need further explanation. As we mentioned, 14 extra qubits are needed to evaluate the 7 $T/T^{\dagger}$-gates. However in figure \ref{circuito_final} there are only two extra qubits denoted by $q_3$ and $q_4$. Instead of using 7 EPR pairs and performing 7 $S^a$-rotated Bell measurements, we make use of Qiskit reset operation which is denoted by the gate $\ket{0}$ in the circuit. This gate simply turns the state back to $\ket{0}$, so we can perform a $S^a$-rotated Bell measurement, use the reset operation on the qubits that were measured and entangle them again using a $H$-gate and a $CNOT$ gate. The reason this is done is simply to obtain a more compact image, because instead of using 17 qubits only 5 are needed. The number of $S^a$-rotated Bell measurements remains the same: $M=7$. 

Notice the classically controlled $S$-gates used in the circuit. These gates are denoted by a $S$-gate on the target qubit and a control located in the register denoted by $c$ in the circuit, below $q_4$. This $c$ register contains the results of all the quantum measurements so it is made of classical bits. An example of these classically controlled $S$-gates are the two gates that act on $q_3$ after the first $T$ (second $T/T^{\dagger}$) gate of the circuit. Recall from section \ref{sec:encryption} that in the decryption process the client needs to obtain the corresponding value of $a_j$ before performing the correct $S^a$-rotated Bell measurement. This means that to perform the correct simulation one cannot simply apply $S^1$-rotated Bell measurements in all cases since there would be cases where no $S$-gate should be applied in the measurement. Instead each possible measurement result from all the previous $S^a$-rotated Bell measurements has to be taken into account and used to calculate the corresponding $a$ for the next measurement so the simulation returns the correct values for all possible scenarios. Using the key updating rules explained previously, the $a_j$ relative to a measurement is calculated, one for each possible result of all the previous measurements. Then for each $a_j=1$ a controlled $S$-gate is used, where the control is a classical bit string sequence stored in $c$ that contains one result of all the previous measurements. As an example notice that for the second $T/T^{\dagger}$ in figure \ref{circuito_final} there are two control-$S$ gates. This is because in the previous $T/T^{\dagger}$-gate four measurement results are obtained. After key updating two of them leads to $a_{17}=1$ for this second $T/T^{\dagger}$-gate (which is the 18th gate of the circuit), so two classical bit strings stored in $c$ have to be used as the control for two control-$S$ gates. The third $T/T^{\dagger}$ in figure \ref{circuito_final} contains 8 control-$S$ gates for the same reasons, only now there are 16 possible measurement results because there are 2 previous measurements. The second reason why the circuit shown in figure \ref{circuito_final} is not exactly the circuit that was simulated is the number of these control-$S$ gates that were used. For the last $T/T^{\dagger}$ of the circuit 2048 control-$S$ gates were used, so the resulting image would be too large to show here. The only two differences between the circuit in figure \ref{circuito_final} and the simulated circuit are that the latter used 17 qubits with no reset operation and that it contained all the necessary control-$S$ gates in the measurements. Nevertheless the circuit in figure \ref{circuito_final} is still useful to be presented here because it illustrates all the relevant elements of the simulation: the encryption using $X$ and $Z$ gates, the 7 $S^a$-rotated Bell measurements and some classically control-$S$ gates that are needed in these measurements.

Now that these details have been properly discussed
we can proceed with the remaining steps. After evaluating all the quantum gates in the circuit using the key updating rules explained, the server would send the encrypted qubits back to the client, the key-updating functions and all the entangled qubits. As a remark recall that the 7 $S^a$-rotated Bell measurements in the protocol are performed by the client once he receives all the data from the server and updates its keys accordingly. The last step is measuring $q_0$, $q_1$ and $q_2$ and using the final key $dk$ to decrypt the results. 

\begin{figure}[]
	\includegraphics[width=0.48\textwidth]{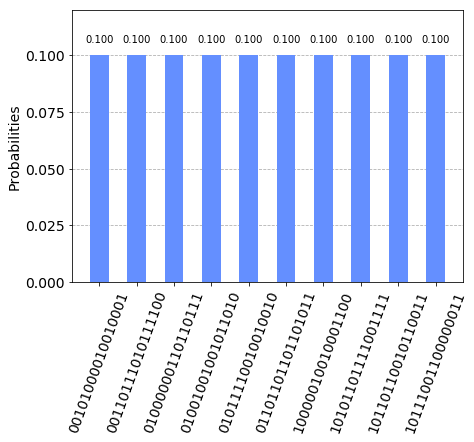}
	\centering
	\caption{Results of the 3 qubits Grover circuit simulation. This histogram shows 10 different measurement results. Each bar contains the results of the 7 $S^a$-rotated Bell measurements and the final encrypted states of qubits $q_2$, $q_1$ and $q_0$.
	}
	\label{resultados}
\end{figure}

The results from the simulation are shown in figure \ref{resultados}. 
The horizontal axis represents the measurement result obtained and vertical axis shows the probability of obtaining each result. The circuit was executed 10 times in order to obtain a compact image. Every state in figure \ref{resultados} is obtained with probability $0.1$\%. Each measurement result contains the results obtained from the 7 $S^a$-rotated Bell measurements and the measurement values of $q_0$, $q_1$ and $q_2$. The first 14 bits in each result represent the results of the $S^a$-rotated Bell measurements. If we start reading from the left, the first two bits refer to the last measurement and the last two refer to the first. This is because the first Bell measurement is performed on qubits $q_4$ and $q_3$. The next $S^a$-rotated Bell measurement is performed on qubits $q_6$ and $q_5$ and so on. Then the last 3 remaining bits represent $q_2$, $q_1$ and $q_0$. The probability of each state of figure \ref{resultados} is not the main focus here. Instead what we are interested in is obtaining the correct result after decryption given a measurement result of $q_0$, $q_1$ and $q_2$. 

\begin{table*}[]
	\begin{tabular}{|c|c|c|c|c|c|c|}
		\hline
		Simulation result & $S^a$-rotated Bell measurements & Encrypted result & Final $q_2$ key & Final $q_1$ key & Final $q_0$ key & Decrypted result \\ \hline
		00101000010010001 & 00101000010010                 & 001              & (0,1)           & (1,1)           & (0,1)           & 011              \\ \hline
		00110111010111100 & 00110111010111                 & 100              & (0,1)           & (0,0)           & (1,0)           & 101              \\ \hline
		01000000110110111 & 01000000110110                 & 111              & (0,0)           & (1,0)           & (0,1)           & 101              \\ \hline
		01001001001011010 & 01001001001011                 & 010              & (0,0)           & (0,0)           & (1,0)           & 011              \\ \hline
		01011110010010010 & 01011110010010                 & 010              & (0,0)           & (0,1)           & (1,1)           & 011              \\ \hline
		01101101101101011 & 01101101101101                 & 011              & (0,1)           & (0,1)           & (0,0)           & 011              \\ \hline
		10000010010001100 & 10000010010001                 & 100              & (1,0)           & (1,0)           & (1,1)           & 011              \\ \hline
		10101101111001111 & 10101101111001                 & 111              & (1,1)           & (0,0)           & (0,0)           & 011              \\ \hline
		10110110010110011 & 10110110010110                 & 011              & (0,1)           & (0,1)           & (0,0)           & 011              \\ \hline
		10111001100000011 & 10111001100000                 & 011              & (1,1)           & (1,0)           & (0,1)           & 101              \\ \hline
	\end{tabular}
	\caption{Results from figure \ref{resultados}. For each result obtained from the simulation the following information is given: its 7 Bell measurements results, the encrypted Grover search result, the corresponding final key for each qubit $q_2$, $q_1$ and $q_0$ and the final Grover search result.}
	\label{tab:my-table}
\end{table*}

As an example, if we take the first result from figure \ref{resultados} which is $00101000010010001$, the last 3 bits $001$ represent the encrypted value of Grover's algorithm result. The state $001$ is not one of the solutions. On the other hand $00101000010010$ are the results of the Bell measurements. Using these values and our key updating script based on the rules from section \ref{sec:encryption} the final key can be obtained. In this case the final key is $\left(a_{\text{final}}(0),b_{\text{final}}(0)  \right)=(0,1)$ for $q_0$, $\left(a_{\text{final}}(1),b_{\text{final}}(1)  \right)=(1,1)$ for $q_1$ and $\left(a_{\text{final}}(2),b_{\text{final}}(2)  \right)=(0,1)$ for $q_2$ so  $dk=(a_{\text{final}},b_{\text{final}})=(a_{\text{final}}(0)a_{\text{final}}(1)a_{\text{final}}(2),b_{\text{final}}(0))b_{\text{final}}(1)b_{\text{final}}(2))=(010,111)$. Finally the result of the algorithm can be unencrypted. Since the measurement of $q_0$, $q_1$ and $q_2$ was performed before decrypting the qubits using the final key, to decrypt a classical result the operation is simply $a_{\text{final}}\oplus r_c$, the bitwise modulo 2 sum where $r_c$ is the classical bit string obtained from the quantum measurement. Therefore for this example where the encrypted result of Grover's algorithm is $r_c=001$ we have $010\oplus001=011$. Recall that for the evaluated Grover circuit the correct results are either $011$ or $101$, so the result that was obtained is indeed correct.

For the remaining 9 results obtained from figure \ref{resultados}, the same process was performed to obtain the decrypted result. Table \ref{tab:my-table} shows the final unencrypted results along the final key of every qubit for each result obtained from the Qiskit simulation. As it can be seen every simulation result is correctly decrypted into $011$ or $101$, the desired elements from the Grover search problem. The bit string $011$ was obtained 7 times and $101$ was obtained the remaining 3. If the simulation is performed more times, the results get closer to the expected 50\% chance of obtaining each state. Therefore the simulation obtains the correct results of the algorithm after decryption for all possible cases, demonstrating the correctness of the QHE scheme.

\section{General homomorphic Grover evaluation}
\label{sec:proposal}
In this section the $T/T^{\dagger}$-gate complexity of Grover's algorithm is discussed, along with a proposal to evaluate any Grover circuit homomorphically. For this purpose, the oracle and the diffusion operator have to be decomposed into gates from the set $\mathcal{G}= \{X,Z,H,S,CNOT,T,T^{\dagger}\}$. We will consider the case where there is only one marked element in the list and then generalize for the case with $m$ marked elements.  

Starting with the oracle $U_w$, a quantum state is marked which is the solution of the problem. It can be constructed using a multi-controlled $Z$-gate with $n-1$ control qubits that will add a negative phase only to the desired quantum state. These control qubits should activate only for the desired state. For each qubit in which $\ket{w}$ contains a 0 the corresponding control qubit has to be surrounded by $X$-gates. This is also true for the target qubit. As we have seen, a controlled $Z$-gate can easily be decomposed into a $CNOT$ by surrounding the target qubit with 2 $H$-gates, so the oracle can be implemented with a multi-controlled $CNOT$ gate with $n-1$ control qubits. 

The diffusion operator on the other hand, is independent of the searched element and is the same for every Grover circuit. Since $U_s=2\ket{s}\bra{s}-I$ and $\ket{s}=H^{\otimes n} \ket{0^{\otimes n}}$ we have:
\begin{equation}
U_s=2\ket{s}\bra{s}-I=H^{\otimes n}\left(2\ket{0^{\otimes n}}\bra{0^{\otimes n}}-I\right)H^{\otimes n}.
\end{equation}
This means that the operator $2\ket{0^{\otimes n}}\bra{0^{\otimes n}}-I$ has to be surrounded by $n$ $H$-gates. If a global $(-1)$ phase is applied we have $I-2\ket{0^{\otimes n}}\bra{0^{\otimes n}}$. This operator is basically another oracle in which the desired state is the $\ket{0^{\otimes n}}$ state since $\left[I-2\ket{0^{\otimes n}}\bra{0^{\otimes n}}\right]\ket{0^{\otimes n}}=-\ket{0^{\otimes n}}$ and $\left[I-2\ket{0^{\otimes n}}\bra{0^{\otimes n}}\right]\ket{x}=\ket{x}$ for any $\ket{x}\neq \ket{0^{\otimes}}$. Then 
it can be implemented using a multi-controlled $Z$-gate with $n-1$ control qubits surrounded by $n$ $X$-gates. Therefore the diffusion operator can be implemented using a multi-control $CNOT$ gate with $n-1$ control qubits and a target qubit surrounded by 2 $H$-gates, where all the $n$ qubits are surrounded by $n$ $X$-gates and $n$ $H$-gates. 

\begin{figure}[]
	\includegraphics[width=0.48\textwidth]{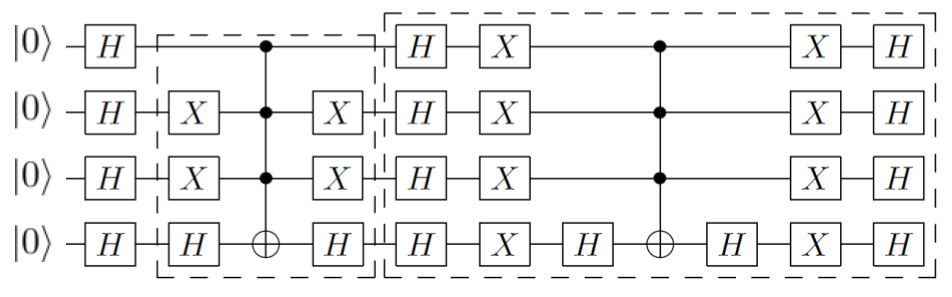}
	\caption{Grover oracle and diffusion operators for $w=1001$ using gates from the set $\mathcal{G}$ and two multi-controlled $CNOT$ gates with 3 control qubits. Both operators are surrounded by a dotted box. The operators are used $O(\sqrt{N})$ times and then each qubit is measured.}
	\label{grover_4}
\end{figure}

The whole algorithm is shown in figure \ref{grover_4} for the state $\ket{1001}$ as an example, to make clear how the oracle and diffusion operator are implemented using gates from the set $\mathcal{G}= \{X,Z,H,S,CNOT,T,T^{\dagger}\}$. The only gates that are not from this set are the multi-controlled $CNOT$ gates with 3 control qubits.

The number of $T/T^{\dagger}$-gates is the relevant parameter that determines if a certain algorithm can be efficiently encrypted using Liang's QHE scheme. Then the only gates that have to be taken into account from the oracle and the diffusion operator are the multi-controlled $CNOT$ gates, since these gates are the only ones that contain $T/T^{\dagger}$-gates.  Recall that the Grover algorithm needs $n$ qubits to search $N=2^n$ elements in a list. For each iteration of the Grover algorithm the oracle and the diffusion operator are applied once. Then in the case there is only one marked state, one iteration of Grover's algorithm requires two multi-controlled $CNOT$ gates (with $n-1$ control qubits), one for the oracle and another for the diffusion operator. The next step is expressing these gates in terms of $T/T^{\dagger}$-gates.

\begin{figure}[]
	\includegraphics[width=0.52\textwidth]{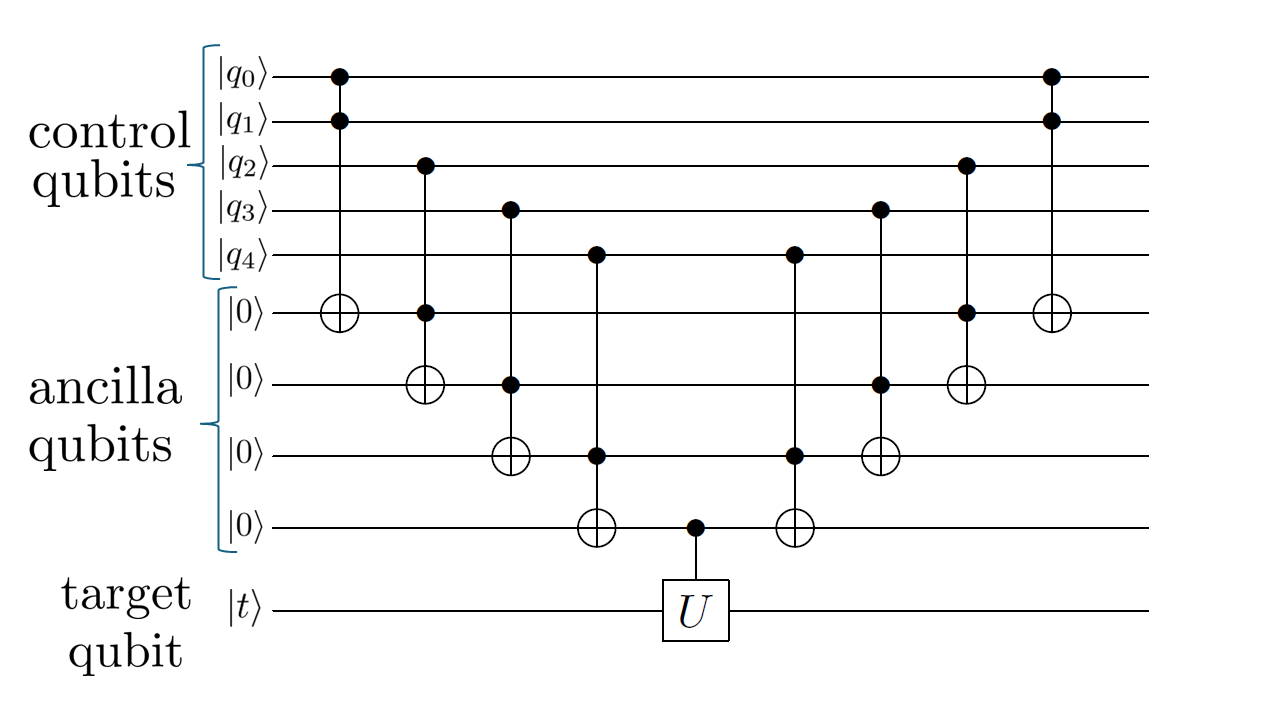}
	\caption{A $n=6$ multi-controlled single qubit gate with 5 control qubits, 4 ancilla qubits and 8 Toffoli gates. }
	\label{chuang cnots}
\end{figure}

A multi-controlled $CNOT$ gate with $n-1$ control qubits can be decomposed into $2(n-2)$ Toffoli gates using $(n-2)$ extra ancilla qubits. An example of this decomposition \cite{Nielsen} is shown in figure \ref{chuang cnots} for any single qubit gate $U$. If $U=X$ this corresponds to the decomposition of the multi-controlled $CNOT$ gate.

If two multi-controlled $CNOT$ gates are used to run one query of the Grover algorithm for a list with just one marked item, then the number of Toffoli gates are given by:

\begin{multline}
2\quad \text{MCNOT}= 2 (2(n-2)) \quad \text{Toffolis}=\\=4(n-2) \quad \text{Toffolis} =4(\log(N)-2) \quad \text{Toffolis}
\label{MultiTof}
\end{multline}
where $n=\log(N)$ was used. Since the oracle and the diffusion operator are repeated $O(\sqrt{N})$ times, the number of Toffoli gates needed to run the whole algorithm is simply given by:
\begin{equation}
4[\log(N)-2)]\cdot\sqrt{N} \quad \text{Toffolis}.
\label{algoritmo}
\end{equation}

Since a Toffoli gate can be decomposed in 7 $T/T^{\dagger}$-gates (figure \ref{Toffolis}) the number of $T/T^{\dagger}$-gates, $M$, is finally obtained:
\begin{equation}
M= 28[\log(N)-2)]\cdot\sqrt{N} \quad \rightarrow O(\log(N)\sqrt{N}).
\label{numberT}
\end{equation}
$M$ grows slower than a linear function. Since the QHE scheme is only suitable for circuits with a polynomial number of $T/T^{\dagger}$-gates, the fact that $M=O(\log(N)\sqrt{N})$ makes the Grover algorithm a protocol that can be implemented efficiently using the QHE scheme discussed. If there are more desired elements in the $N$ items list instead of just 1, this result is still true. The reason is that for each new marked element, the diffusion operator stays the same and only the oracle changes. Instead of just using 1 multi-controlled $CNOT$ gate, the oracle would be implemented using $m$ multi-controlled $CNOT$ gates. Then the number of $T/T^{\dagger}$-gates is still $O(\log(N)\sqrt{N})$ and so the decryption process of the algorithm remains efficient.
Therefore the Grover algorithm can be homomorphically implemented using $n$ qubits and $n-2$ extra ancilla qubits in an efficient manner. For a $N=2^n$ list, the client should then prepare $2n-2$ qubits in total, encrypt them, send them to the server and then perform the corresponding $M$ measurements to correctly decrypt the results.

To conclude this section we want to mention some results regarding the special case in which there is only one desired element in the database. Grover \cite{Grover 2002} proposed an algorithm that could find this unique item using only $O(\sqrt{N}\log \log N)$ elementary gates without increasing the number of queries needed significantly. This algorithm is no longer made of $O(\sqrt{N})$ identical iterations and it is more complicated than the usual Grover algorithm. Then in 2017 Arunachalam and de Wolf \cite{Grover optimization} proposed an even more efficient Grover search algorithm regarding its gate complexity. For a sufficiently large database of size $N$ that contains just one desired element and for any constant $r$ this algorithm finds the only solution using $O(\sqrt{N})$ queries and $O(\sqrt{N}\log^{(r)}N)$ elementary gates. The iterated binary logarithm is defined as $\log^{(s+1)}=\log \circ \log^{(s)}$ where $\log^{(0)}$ is the identity function. The elementary gates allowed are the Toffoli gate and any unitary single qubit gate like the $H$ and $X$ gates. If we have a very large $N$ items list that is also a power of 2, $r$ can be chosen to be $r=\log^{\star} N$ so the algorithm finds the searched element using only $O(\sqrt{N}\log(\log^{\star}N))$ elementary gates in the optimal $\frac{\pi}{4}\sqrt{N}$ queries. Here the function $\log^{\star}N$ represents the number of times the binary logarithm must be iteratively applied to $N$ to obtain a number that is at most 1: $\log^{\star}N=\text{min}\{r\geq0: \log^{(r)}N \leq1\}$. Since these algorithms have a reduced number of $T/T^{\dagger}$-gates compared to the usual Grover algorithm, they can also be homomorphically evaluated  using the QHE scheme discussed. Furthermore, their homomorphic evaluation is even more efficient than the usual Grover algorithm due to their reduced $T/T^{\dagger}$-gate complexity. 

Arunachalam and de Wolf \cite{Grover optimization} mentioned that most applications of Grover's algorithm study databases with an unknown number of desired elements and only focus on the number of queries. They ended asking whether there are any applications where the reduction in the number of quantum gates for the special case of just one marked element is both applicable and significant. We want to answer this question positively. 
Quantum homomorphic encryption is a perfect application for this algorithm, since the advantage it offers regarding the gate complexity is significant in the context of improving the efficiency of the decryption process. Therefore, to solve the quantum search problem for the particular case of one solution in the database, the most efficient algorithm that can be implemented homomorphically is the one proposed by Arunachalam and de Wolf.

\section{Conclusions}
\label{sec:conclusions}
In this article Liang's \cite{Liang} quasi-compact Quantum Homomorphic Encryption scheme and Grover's algorithm have been combined. This QHE scheme has perfect security, $\mathcal{F}$-homomorphism, no interaction between client and server and quasi-compactness. The scheme is not compact so it does not contradict the no-go result given by Yu \cite{no go result}. The decryption procedure is independent of the size of the evaluated circuit and depends only on the number of $T/T^{\dagger}$-gates contained in the circuit. 

Making use of this QHE scheme, a quantum circuit that implements the Grover algorithm homomorphically for a particular case has been simulated using Qiskit. We showed how to apply controlled-$S$ gates in the Qiskit simulation in order to account for every possible result. The results obtained from this simulation can always be correctly decrypted after obtaining the final key for each qubit involved.  

Regarding the complexity of the decryption procedure of the QHE scheme, it is only efficient for circuits with a polynomial number of $T/T^{\dagger}$-gates. As it has been seen, the number of $T/T^{\dagger}$-gates of any Grover circuit is $O(\log(N)\sqrt{N})$ which means it grows slower than any linear function. 
Therefore, Grover's algorithm is a perfect example of a quantum algorithm that can be evaluated homomorphically with perfect security and non interaction in an efficient manner. In the particular case of a database with just one solution, a more efficient algorithm proposed by Arunachalam and de Wolf can be used. Its gate complexity is $O(\sqrt{N}\log(\log^{\star}N))$ so the decryption process of its homomorphic evaluation is more efficient than the usual Grover algorithm.

Future works in this area of research could be studying the $T/T^{\dagger}$-gate complexity of more quantum algorithms, in order to analyse their homomorphic implementation using Liang's QHE scheme. In the case it is not possible to implement them homomorphically in the most general scenario, there may be particular cases, similar to Grover's search on a database with just one marked element, where the $T/T^{\dagger}$-gate complexity is low enough so the homomorphic implementation is still efficient. These restricted algorithms would then still be useful as an application of quantum homomorphic encryption.

\section{Acknowledgements}
We acknowledge support from grants MINECO/FEDER Projects, PID2021-122547NB-I00 FIS2021, the ``MADQuantumCM'' project funded by Comunidad de Madrid. M.A.M.-D. has been financially supported by the Ministry of Economic Affairs and Digital Transformation of the Spanish Government through the QUANTUM ENIA project call-Quantum Spain project, and by the European Union through the Recovery, Transformation and Resilience Plan-NextGenerationEU within the framework of the Digital Spain 2026 Agenda. M.A.M.-D. has been partially supported by the U.S.Army Research Office through Grant No. W911NF-14-1-0103. P.F.O. acknowledges support from
a MICINN contract PRE2019-090517 (MICINN/AEI/FSE,UE).

\section*{Declarations}
\textbf{Conflict of Interest} The authors declare that they have no competing interests that are relevant to the content of this article.

\end{document}